# A New Sampling Method Base on Sequential Tests with Fixed Sample Size Upper Limit

Dihong Huang*
Shanghai Jiao Tong University
Shanghai, China
huangdihong@sjtu.edu.cn

*Abstract*—Sequential inspection is a technique employed to monitor product quality during the production process. For smaller batch sizes, the Acceptable Quality Limit (AQL) inspection theory is typically applied, whereas for larger batch sizes, the Poisson distribution is commonly utilized to determine the sample size and rejection thresholds. However, due to the fact that the rate of defective products is usually low in actual production, using these methods often requires more samples to draw conclusions, resulting in higher inspection time. Based on this, this paper proposes a sequential inspection method with a fixed upper limit of sample size. This approach not only incorporates the Poisson distribution algorithm, allowing for rapid calculation of sample size and rejection thresholds to facilitate planning, but also adapts the concept of sequential inspection to dynamically modify the sampling plan and decision-making process. This method aims to decrease the number of samples required while preserving the inspection's efficacy. Finally, this paper shows through Monte Carlo simulation that compared with the traditional Poisson distribution algorithm, the sequential test method with a fixed sample size upper limit significantly reduces the number of samples compared to the traditional Poisson distribution algorithm, while maintaining effective inspection outcomes.

## I. INTRODUCTION

Product testing is an important method used by companies to determine whether the finished product meets quality requirements during the product manufacturing process. However, it is not possible to test all products due to the destructive and irreversible nature of some testing methods, i.e. the finished product will be destroyed after being tested. Sampling testing method is a product testing method often adopted by enterprises, which takes a certain number of products in the same batch for inspection and determines whether the products in this batch meet the quality requirements by setting a reasonable rejection threshold, i.e., whether the number of defective products in the sampled finished products reaches this threshold. In the research of sampling testing methods, how to scientifically determine the number of samples and the rejection threshold is an important research topic.

In the research of sampling and testing methods, binomial and Poisson distributions are widely used to determine the number of samples to be taken and the rejection threshold. When the number of batches is large and the defective rate is low, the Poisson distribution can be used to calculate the number of spare parts to be sampled, and this is used as the critical condition for accepting and rejecting spare parts [1].

For sampling and testing of large batches, hypothesis testing methods with binomial and normal distributions are also widely used, especially in assessing whether the defective rate is in accordance with the nominal value at different confidence levels [2][3]. Acceptable Quality Level (AQL) is another sampling and testing criterion commonly used in small lot sizes, which is based on statistical principles to predict the overall quality level from a small number of samples [4]. In addition, hypergeometric distributions have also been used to determine sampling inspection criteria, especially in small lot sizes, and this method provides a more accurate estimate of the number of nonconforming products [5]. Traditional sampling and testing methods usually involve first determining the sample size and then testing the products for defective products based on the samples taken in sequence. The inspection is terminated if the number of defective products detected reaches a threshold value, and vice versa, all samples continue to be tested. This type of method may lead to an excessive number of tests when the defective rate is small, thus increasing the inspection time of the enterprise.

In this paper, a sequential inspection method with a fixed sample size upper limit is proposed. First, the method combines the AQL inspection theory [6] and Poisson distribution algorithm to calculate the appropriate sampling and inspection quantity and rejection threshold for the case of low defective rate. In addition, the method introduces the idea of sequential inspection, which dynamically adjusts the sampling quantity by inspecting the products one by one with the help of calculating the decision boundary and likelihood ratio function, so as to terminate the sampling process in advance when the conditions are met [6][7]. Finally, this paper utilizes the Monte Carlo simulation method based on hypergeometric distribution [8][9] to simulate the product inspection process, analyzes the relationship between the sampling quantity, the rejection threshold and the allowable error [10][11], and compares the traditional method with the sequential inspection method with a fixed upper limit of sample size proposed in this paper [12]. The results show that the method proposed in this paper can significantly reduce the number of samples and lower the testing workload in the inspection process.

## II. METHODOLOGY

Fixed sample size upper limit of the sequential test method first according to the batch size to determine the upper limit of the number of extracted. When the sample size is very large, you can use the Poisson distribution to find the number of spare parts to be taken, and the number of unqualified spare parts in the sampling as a critical condition for accepting and rejecting spare parts. The significance level α, the permissible error and defective rate into the Poisson distribution formula, the need to extract the maximum number of spare parts, followed by the reliability and the need to detect the maximum number of spare parts into the Poisson Cumulative Distribution Function (CDF), you can get the threshold for rejection of spare parts, according to which the development of sampling and testing programs. When the batch size is small, the data distribution does not conform to the Poisson distribution, in order to achieve the optimization of the model, we introduce the AQL (Acceptable Quality Level) as the sampling and inspection standard for small batch size. The value of AQL is determined based on the nominal value claimed by the supplier, and the corresponding sampling quantity size is found using the MIL-STD-105E Sampling Inspection Criteria table shown in Fig.1, where Ac stands for the acceptance number and Re stands for the rejection number. Also, for optimization, a sequential inspection scheme is further introduced to reduce the number of samples by calculating likelihood ratios and decision intervals to end the inspection process earlier.

| Sample size code letter | Sample size | Acceptable Quality Levels (normal inspection) | | | | | | | | | | | | | | | | |
|---|---|---|---|---|---|---|---|---|---|---|---|---|---|---|---|---|---|---|
| | | 0.010 | 0.015 | 0.025 | 0.040 | 0.065 | 0.10 | 0.15 | 0.25 | 0.40 | 0.65 | 1.0 | 1.5 | 2.5 | 4.0 | 6.5 | 10 | 15 | 25 |
| | | Ac Re | Ac Re | Ac Re | Ac Re | Ac Re | Ac Re | Ac Re | Ac Re | Ac Re | Ac Re | Ac Re | Ac Re | Ac Re | Ac Re | Ac Re | Ac Re | Ac Re | Ac Re |
| A | 2 | ↓ | ↓ | ↓ | ↓ | ↓ | ↓ | ↓ | ↓ | ↓ | ↓ | ↓ | ↓ | ↓ | 0 1 | ↓ | ↓ | ↓ | 1 2 |
| B | 3 | | | | | | | | | | | | | 0 1 | ↑ | ↓ | 1 2 | 2 3 |
| C | 5 | | | | | | | | | | | | 0 1 | ↑ | ↓ | 1 2 | 2 3 | 3 4 |
| D | 8 | | | | | | | | | | | 0 1 | ↑ | ↓ | 1 2 | 2 3 | 3 4 | 5 6 |
| E | 13 | | | | | | | | | | 0 1 | ↑ | ↓ | 1 2 | 2 3 | 3 4 | 5 6 | 7 8 |
| F | 20 | | | | | | | | | 0 1 | ↑ | ↓ | 1 2 | 2 3 | 3 4 | 5 6 | 7 8 | 10 11 |
| G | 32 | | | | | | | | 0 1 | ↑ | ↓ | 1 2 | 2 3 | 3 4 | 5 6 | 7 8 | 10 11 | 14 15 |
| H | 50 | | | | | | | 0 1 | ↑ | ↓ | 1 2 | 2 3 | 3 4 | 5 6 | 7 8 | 10 11 | 14 15 | 21 22 |
| J | 80 | | | | | | 0 1 | ↑ | ↓ | 1 2 | 2 3 | 3 4 | 5 6 | 7 8 | 10 11 | 14 15 | 21 22 | ↑ |
| K | 125 | | | | | 0 1 | ↑ | ↓ | 1 2 | 2 3 | 3 4 | 5 6 | 7 8 | 10 11 | 14 15 | 21 22 | ↑ | |
| L | 200 | | | | 0 1 | ↑ | ↓ | 1 2 | 2 3 | 3 4 | 5 6 | 7 8 | 10 11 | 14 15 | 21 22 | ↑ | | |
| M | 315 | | | 0 1 | ↑ | ↓ | 1 2 | 2 3 | 3 4 | 5 6 | 7 8 | 10 11 | 14 15 | 21 22 | ↑ | | | |
| N | 500 | | 0 1 | ↑ | ↓ | 1 2 | 2 3 | 3 4 | 5 6 | 7 8 | 10 11 | 14 15 | 21 22 | ↑ | | | | |
| P | 800 | 0 1 | ↑ | ↓ | 1 2 | 2 3 | 3 4 | 5 6 | 7 8 | 10 11 | 14 15 | 21 22 | ↑ | | | | | |
| Q | 1250 | 0 1 | ↑ | 1 2 | 2 3 | 3 4 | 5 6 | 7 8 | 10 11 | 14 15 | 21 22 | ↑ | | | | | | |
| R | 2000 | ↑ | | 1 2 | 2 3 | 3 4 | 5 6 | 7 8 | 10 11 | 14 15 | 21 22 | ↑ | | | | | | |

Fig. 1. MIL-STD-105E Sampling Standards

The Poisson distribution is suitable for describing the probability distribution of the number of occurrences of a random event per unit time. In a Bernoulli test with binomial distribution, the probability of occurrence of an event can be approximated by the Poisson distribution if the number of trials $n$ is large, the probability $p$ of the binomial distribution is small, and the product $\lambda = np$ is moderate. Next, the upper limit of the number of samples to be sampled for testing is calculated. Specifically, the expression for the number of samples sampled for testing $n$ is:

$$n = \frac{Z_{\frac{\alpha}{2}}^2 \cdot p_0(1 - p_0)}{\delta^2} \quad (1)$$

where $\alpha$ is the producer's risk, determined by the confidence level, $p_0$ is the supplier's nominal defective rate, $Z_{\frac{\alpha}{2}}$ is the absolute value of the standard normal distribution, determined by the confidence level, δ is the allowable error rate.

Then, the original and opposing hypotheses are tested using a one-sided Poisson test. Specifically, let the original hypothesis $H_0$: the defective rate of spare parts does not exceed the nominal value; and the opposing hypothesis $H_1$: the defective rate of spare parts exceeds the nominal value. Thresholds were calculated using the Poisson cumulative distribution function (CDF):

$$P(x \leq k_0) = \sum_{i=0}^{k_0} \frac{e^\lambda \lambda^i}{i!} \quad (2)$$

where $k_0$ is the actual number of substandard products, $i$ is the number of substandard products observed in the sample testing, $\lambda$ is the mathematical expectation of the number of times a random event occurs.

However, when the batch size of the parts to be inspected by the enterprise is small, it does not conform to the Poisson distribution, and then the sampling strategy using the Poisson distribution model will produce a large error. Therefore, in order to optimize our sampling strategy, we use the AQL (Acceptable Quality Level) method based on quality sampling theory for sampling when the batch size is small, AQL is a key metric in quality control, it is the maximum number of defective items that can be considered acceptable in random sampling of a production lot for inspection. This paper will determine the number of samples, AQL size, for each nominal value and for the total number of samples according to the MIL-STD-105E sampling standard. When the number of defective samples is less than Ac, the sample is accepted; when the number of defective samples is greater than Re, the sample is rejected. Next, we propose to use a sequential testing scheme with a fixed upper sample limit. Specifically, as a dynamic decision-making method, unlike traditional sampling, SPRT decides the hypothesis by sequentially testing the current observations, i.e., the number of samples is not determined in advance, and the samples are taken one by one, and then stopped when the number of substandard samples obtained is sufficient to make a decision on the given conditions. The specific steps are as follows:

*1) Propose a hypothesis*

- Null hypothesis $H_0$: The defective rate of spare parts is equal to or does not exceed the nominal value, i.e., $p \leq \lambda_0$.
- Alternative Hypothesis $H_1$: The reject rate of spare parts exceeds the nominal value, i.e., $p > \lambda_0$. p represents the actual reject rate of spare parts.

*2) Determine the decision boundary and likelihood function*

First, the bound for rejecting $H_0$ is $B = \frac{1-\beta}{\alpha}$ and the bound for accepting $H_0$ is $A = \frac{\beta}{1-\alpha}$, where $\alpha$ is the maximum allowable probability of making a Type I error and $\beta$ is the maximum allowable probability of making a Type II error. The likelihood ratio function is then defined as follows:

$$\Lambda = \frac{p_1^x(1-p_1)^{n-x}}{p_0^x(1-p_0)^{n-x}} \qquad (3)$$

where $p_0$ is the probability of success of $H_0$ under the null hypothesis, $p_1$ is the probability of success of $H_1$ under the alternative hypothesis, $n$ is the number of samples taken and tested, $x$ denotes the number of defective products in the detection process.

*3) Determining stopping rules*

- If $\Lambda \leq A$, accept the original hypothesis $H_0$;
- If $\Lambda \geq B$, reject the original hypothesis $H_0$;
- If $A < \Lambda < B$ and $n < n_{max}$, continue sampling;
- If $A < \Lambda < B$ and $n = n_{max}$, stop sampling;
- If $x = k^*$, stop sampling.

TABLE I. NUMBER OF SAMPLES FOR TESTING IN TWO CASES

| Case | $\alpha$ | $p_0$ | $Z_{\frac{\alpha}{2}}$ | $\delta$ | n |
|---|---|---|---|---|---|
| Case I | 0.05 | 0.1 | 1.96 | 0.05 | 139 |
| Cse II | 0.1 | 0.1 | 1.645 | 0.05 | 98 |

TABLE II. THE THRESHOLD VALUE IN TWO CASES

| Case | Threshold Type | Conditions | Threshold $k^*$ |
|---|---|---|---|
| Case I | rejection | $P(x \geq k*|H_0) \leq 0.05$ | 21 |
| Case II | acceptance | $P(x \leq k*|H_0) \geq 0.90$ | 15 |

TABLE III. SAMPLING AND TESTING METHODS OF TWO CASES

| Case | Sample Size | Decisions |
|---|---|---|
| Case I | 139 | If a sample has more than 21 non-conforming products, the parts are rejected. |
| Case II | 98 | If the number of non-conforming products tested in a sample is less than 15, the parts will be accepted. |

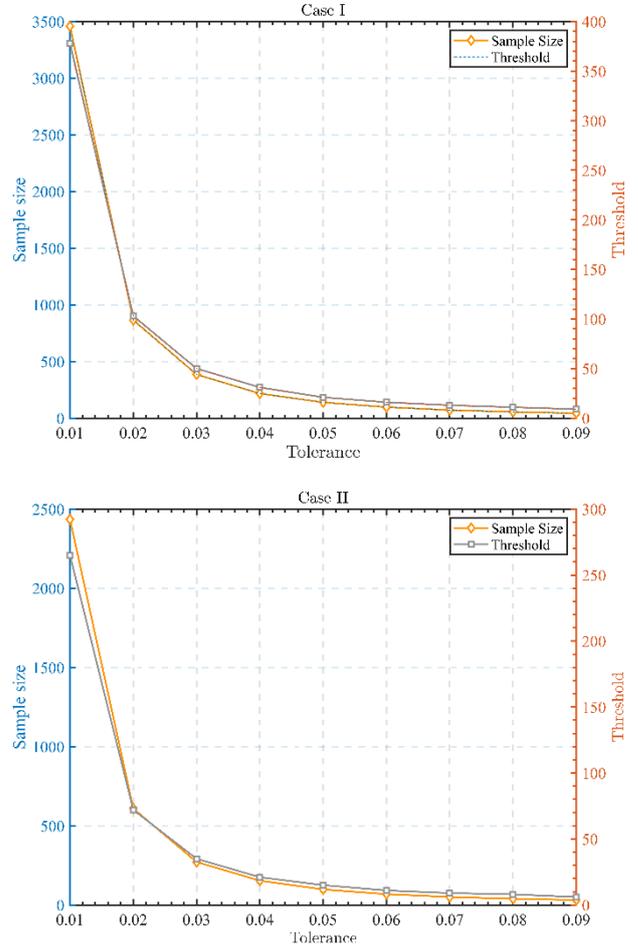

Fig. 2. Relationship between sample size, threshold and tolerance

## III. EXPERIMENTS

First, in order to verify that the sequential test method with fixed sample size can effectively reduce the number of samples, this paper selected two common scenarios to analyze the experimental data. The specific experimental settings of two cases are shown in TABLE I, N represents the number of samples.

Next, we can compute the probability distribution of the number $k^*$ of non-confirming products for Case I and Case I under $H_0$, the result of which is shown in TABLE II.

Then, the relationship between the number of samples sampled, the threshold value, and the for C allowable error Case I and Case I is shown in Fig. 2.

As can be seen from Fig. 2, the sample size and the threshold decrease as the allowable error level increases. To satisfy the allowable error $\delta \leq 0.05$ and to keep the number of sampling tests as small as possible, scheme should be used is shown in TABLE III.

TABLE IV. SAMPLING AND TESTING METHOD FOR CASE I

| Batch size | AQL | | | |
|---|---|---|---|---|
| | 0.025 | 0.04 | 0.065 | 0.1 |
| 2-8 | (2,0) | (2,0) | (2,0) | (2,0) |
| 9-15 | (3,0) | (3,0) | (3,0) | (3,0) |
| 16-25 | (5,0) | (5,0) | (5,0) | (5,1) |
| 26-50 | (8,0) | (8,0) | (8,1) | (8,2) |
| 51-90 | (13,0) | (13,1) | (13,2) | (13,3) |
| 91-150 | (20,1) | (20,2) | (20,3) | (20,5) |
| 151-280 | (32,2) | (32,3) | (32,5) | (32,7) |
| 281-500 | (38,8) | (50,5) | (50,7) | (50,10) |
| 501-1200 | | (60,11) | (80,10) | (80,14) |
| 1201-3200 | | | (94,16) | (125,21) |
| >3200 | | | | (139,21) |

TABLE V. SAMPLING AND TESTING METHOD FOR CASE II

| Batch size | AQL | | | |
|---|---|---|---|---|
| | 0.025 | 0.04 | 0.065 | 0.1 |
| 2-8 | (2,0) | (2,0) | (2,0) | (2,0) |
| 9-15 | (3,0) | (3,0) | (3,0) | (3,0) |
| 16-25 | (5,0) | (5,0) | (5,0) | (5,1) |
| 26-50 | (8,0) | (8,0) | (8,1) | (8,2) |
| 51-90 | (13,0) | (13,1) | (13,2) | (13,3) |
| 91-150 | (20,1) | (20,2) | (20,3) | (20,5) |
| 151-280 | (27,6) | (32,3) | (32,5) | (32,7) |
| 281-500 | | (42,8) | (50,7) | (50,10) |
| 501-1200 | | | (66,11) | (80,14) |
| 1201-3200 | | | | (98,15) |
| >3200 | | | | |

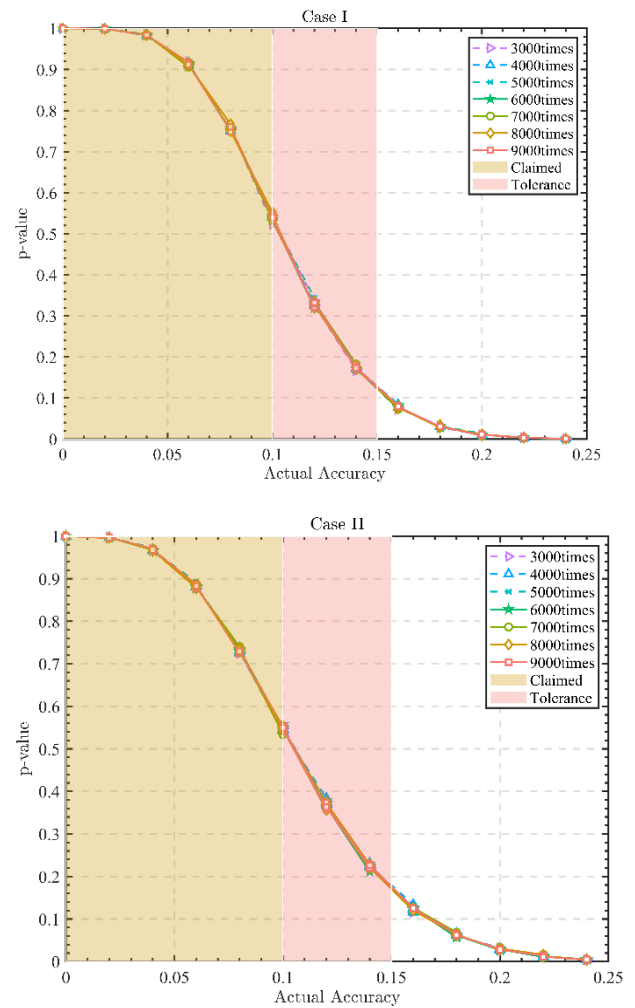

Fig. 3. Results of Monta Carlo simulations

Next, to verify whether it is accurate to regard the defective rate as satisfying the Poisson distribution when the batch size is large, this paper run Monte Carlo Simulation to simulate the sampling process. At the same time, because the way of drawing samples is to draw all the samples to be tested at the same time, this paper will base on the hypergeometric distribution to construct Monte Carlo Simulation to test the accuracy of the model. We basically performed Monte Carlo Simulation for 3000, 4000, 5000, 6000, 7000, 8000, 9000 times, and estimated the relation between p-value and the actual accuracy. During the experiments, in Case I, we tested 139 products, while in Case II we tested 98 products. We draw the relationship between the level of significance and the actual accuracy, as shown in Fig. 3.

As can be seen in Fig. 3, the six curves are basically overlapped in Case I and Case II, which shows that the algorithm using Poisson distribution in this paper has high accuracy as well as high robustness. Then, in order to enable the enterprises clearly and conveniently use the sampling test method, this paper combines the Poisson distribution and AQL quality testing theory, with a table listing the sampling test program corresponding to different sample sizes, specific as shown in TABLE IV, TABLE V. The meaning of (n, c) in the table is (the number of sampling test, the threshold of unqualified products). For example, if the company's tolerance level is $\delta = 0.05$ and the producer claims that the non-confirming rate is $p_0 = 0.01$, which is identical to Case I, the company can simply use TABLE IV to decide their sampling method. More specifically, when the company get a batch size of 100, they should test 20 products and when more than 4 of them were non-confirming products, the company should reject the whole batch because it is very likely that its non-confirming rate is over 0.01. For Case II, if the company's tolerance level is $\delta = 0.05$ and the company expected that the non-confirming rate $p_0$ is less than 0.025, which is identical to Case II, the company can simply use TABLE V to decide their sampling method. More specifically, when the company get a batch size of 5000, they should test 27 products and when more than 5 of them were non-confirming products, the company should reject the whole batch because it is very likely that this batch can not meet the company's expectation.

In order to test whether the sequential test with a fixed upper sample limit can effectively reduce the number of sampling tests, we set the probability of committing the first type of error $\alpha = 0.05$, the probability of committing the second type of error $\beta = 0.05$, and the true defective rate of the product $p_0 = 0.10$. 10,000 simulated sampling tests were carried out by using the Monte Carlo simulation, and the number of sampling tests was obtained according to the above rule. We compared the difference in the number of sampling tests in the same situation with and without the sequential test method, and described the difference between the two sets of data using the t-test, and the results are shown in Fig. 4.

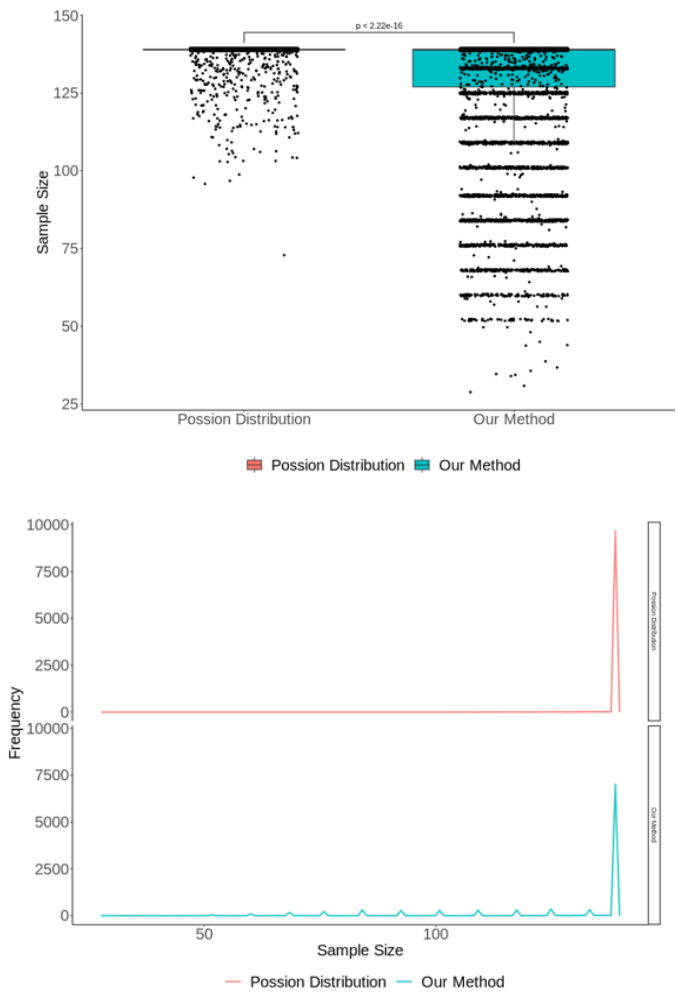

Fig. 4.  Comparison of model performance

As can be seen from Fig. 4, compared with sampling inspection without sequential testing, sequential testing with a fixed upper limit of sample size can effectively reduce the number of sampling inspection. As in the sampling and testing process, the general situation of the product non-confirming rate is low, belonging to the small probability of events, the ordinary sampling and testing methods defective number of cases to reach the threshold is relatively small. If the sequential test method is not used, in most cases it is necessary to reach the upper limit of the sample size before stopping the sampling test. This conclusion can also be verified from the frequency distribution graph. The sequential test with fixed sample size upper limit can effectively combine the sampling upper limit of Poisson distribution and the advantages of dynamic adjustment of the sampling number of sequential probability ratio test, which makes it possible to end the sampling and testing process earlier in some cases and reduces the inspection time.

## IV. Conclusion

This study establishes a sequential test method with a predetermined maximum sample size, which successfully integrates the Poisson distribution algorithm with AQL sampling principles to determine the sample size ceiling and rejection thresholds. By incorporating sequential testing, the number of inspections is minimized, thereby streamlining the inspection process. The experimental findings indicate that the sequential test with a fixed sample size upper limit effectively harnesses the benefits of the Poisson distribution for sample size limits and the sequential probability ratio test for dynamically adjusting the sample size. This allows for the potential to conclude the sampling and testing process prematurely in certain scenarios, thereby optimizing the inspection procedure. Additionally, the research reveals that the sequential test with a maximum sample size often concludes before reaching the predetermined sample size, with the prematurely terminated tests exhibiting multiple minor peaks and displaying a discernible pattern. These minor peaks frequently occur near the sample size ceiling, with various wavelets approximately equidistant from one another. Future research could investigate the underlying causes for the emergence of these peaks and refine the sampling methodology based on our findings.

## References

[1] Ghojogh B, Nekoei H, Ghojogh A, et al. Sampling algorithms, from survey sampling to Monte Carlo methods: Tutorial and literature review[J]. arXiv preprint arXiv:2011.00901, 2020.

[2] Hafeez, Waqar, Nazrina Aziz, and Javid Shabbir. "Estimation of AQL, LQL and Quality Regions for Group Chain Sampling Plan with Binomial Distribution." Thailand Statistician 22, no. 3 (June 29, 2024): 674–687.

[3] Cai J, Feng Q, Horrace W C, et al. Wrong skewness and finite sample correction in the normal-half normal stochastic frontier model[J]. Empirical Economics, 2021, 60(6): 2837-2866.

[4] Ershadi, Mohammad Javad. "Statistical Design of a Sampling Method in Quality Control of Research Data." Iranian Journal of Information Management 8, no. 1 (August 23, 2022): 1–22.

[5] Johannssen, Arne, Nataliya Chukhrova, and Philippe Castagliola. "Efficient Algorithms for Calculating the Probability Distribution of the Sum of Hypergeometric-Distributed Random Variables." MethodsX 8 (2021): 101507.

[6] Giner-Sorolla R, Montoya A K, Reifman A, et al. Power to detect what? Considerations for planning and evaluating sample size[J]. Personality and Social Psychology Review, 2024, 28(3): 276-301.

[7] Ji Q, Zhang D, Zhao Y. Searching for safe-haven assets during the COVID-19 pandemic[J]. International Review of Financial Analysis, 2020, 71: 101526.

[8] Hutchison W D. Sequential sampling to determine population density[J]. Handbook of sampling methods for arthropods in agriculture, 2020: 207-243.

[9] Lakens, Daniël. "Sample Size Justification." Edited by Don van Ravenzwaaij. Collabra: Psychology 8, no. 1 (March 22, 2022): 33267.

[10] Landau, David, and Kurt Binder. A Guide to Monte Carlo Simulations in Statistical Physics. Cambridge University Press, 2021.

[11] Sarrut D, Bała M, Bardiès M, et al. Advanced Monte Carlo simulations of emission tomography imaging systems with GATE[J]. Physics in Medicine & Biology, 2021, 66(10): 10TR03.

[12] Sharma D K, Singh B, Raja M, et al. An Efficient Python Approach for Simulation of Poisson Distribution[C]//2021 7th International Conference on Advanced Computing and Communication Systems (ICACCS). IEEE, 2021, 1: 2011-2014.